\documentstyle[12pt,axodraw]{article}
\begin{document}
\baselineskip 13pt
\begin{center}
\Large
{\bf Next-to-leading virtual photon - Reggeized gluon interaction\footnote{Work supported in part by INTAS and in part
by the Russian Fund of Basic Researches.}}
\end{center}
\vskip 0.2cm
\centerline{  V. Fadin$^{a, b}$, D. Ivanov$^c$ and \underline{M. Kotsky}$^{a, d}$\footnote{The speaker thanks
the Organizing Committee for the very interesting and well organized conference.}}
\vskip 0.2cm
\centerline{\sl $^a$ Budker Institute for Nuclear Physics, 630090 Novosibirsk, Russia}
\centerline{\sl $^b$ Novosibirsk State University, 630090 Novosibirsk, Russia}
\centerline{\sl $^c$ Institute of Mathematics, 630090 Novosibirsk, Russia}
\centerline{\sl $^d$ Istituto Nazionale di Fisica Nucleare, Gruppo collegato}
\centerline{\sl di Cosenza, Arcavacata di Rende, I-87036 Cosenza, Italy}
\vskip 0.2cm
\begin{abstract}
We present  the results of the calculation of the one-loop correction to the
effective vertex for the quark-antiquark pair production in collisions of the
virtual photon with the Reggeized gluon. This vertex is supposed then to be used for the
calculation of the virtual photon  impact factor, which is extremely important, for instance,  for the description of
the small $x$ deep inelastic scattering
in the BFKL approach.
\end{abstract}
\vskip 0.2cm

The vertex $\Gamma^c_{\gamma^*q\bar q}$ we are interested in can be obtained from the projected on the colour
octet state and negative signature in the $t$-channel
amplitude  of the quark-antiquark production in the virtual photon fragmentation region at collisions of this photon
with any particle.
For simplicity we use the amplitude for the collision of a
virtual photon $\gamma^*$ with momentum $p_A$ with a massless quark $Q$ with momentum $p_B$.
The Sudakov decompositions  for the virtual photon polarization vector $e$, the produced quark momentum
$k_1$ and the antiquark one $k_2$ are
\begin{equation}\label{1}
e = e_\perp + \frac{2Q}{s}p_2,\ \ k_1 = xp_1 + \frac{\vec k_1^{~2}}{sx}p_2 + {k_1}_\perp,\ \  k_2 = (1-x)p_1 +
\frac
{\vec k_2^{~2}}{s(1-x)}p_2 + {k_2}_\perp,\ \ k_1^2 = k_2^2 = 0,
\end{equation}
where
\begin{equation}\label{2}
p_A = p_1 - \frac{Q^2}{s}p_2,\ \ p_B = p_2,\ \ p_1^2 = p_2^2 = 0,\ \ s = 2p_1p_2
\rightarrow \infty,\ \ \vec p^{~2} = - p_\perp^2.
\end{equation}
The quark-antiquark pair is produced with $x \sim 1 - x \sim 1$ and limited (not growing with  $s$) transverse
momenta. The virtual
photon is taken in the light-cone gauge transverse to the vector $p_2$. The terms in the amplitude proportional to
$e_\perp$ are responsible for the transverse photon  and the terms proportional to $Q$
are related to the scalar
photon.
We split the one-loop correction $\Gamma_{\gamma^*q\bar q}^{(1)}$ into three parts according to the three types of
the diagrams contributing
to the amplitude ${\cal A}_{Q\gamma^* \rightarrow Qq\bar q}^{(8,-)}$ with colour octet state and
negative signature in the $t$-channel: two-gluon exchange, one-gluon exchange and
$t$-channel gluon self-energy diagrams. The first of these contributions to the vertex can be found from the relation
\begin{equation}\label{3}
\Gamma_{\gamma^*q\bar q}^{(2g)c(1)}\frac{2s}{t}\Gamma_{QQ}^{c(0)} + \Gamma_{\gamma^*q\bar q}^{c(0)}
\frac{2s}{t}\Gamma_{QQ}^{(2g)c(1)} = {\cal A}_{Q\gamma^* \rightarrow Qq\bar q}^{(2g)(8,-)(1)} - \Gamma_
{\gamma^*q\bar q}^{c(0)}\frac{s}{t}\omega^{(1)}(t)\biggl[ \ln\left( \frac{s}{-t} \right) +  \ln\left( \frac{-s}{-t}
\right) \biggr]\Gamma_{QQ}^{c(0)},
\end{equation}
where $t = q^2 \approx - \vec q^{~2}$ is the squared momentum transfer, $\omega^{(1)}$ is the one-loop
Reggeized gluon trajectory and the correction ${\cal A}_{Q\gamma^* \rightarrow Qq\bar q}^{(2g)(8,-)(1)}$ is given
by
the diagrams of Fig.1
\begin{figure}[tb]
\begin{center}
\begin{picture}(400,80)(0,-40)

\ArrowLine(0,0)(50,0)
\ArrowLine(50,0)(100,0)
\ArrowLine(30,50)(100,70)
\ArrowLine(100,30)(30,50)
\Photon(0,50)(27,50){3}{4}
\ArrowLine(27,50)(30,50)
\Gluon(40,0)(80,64.29){2}{12}
\Gluon(80,0)(50,55.71){2}{10}

\Text(0,60)[l]{$p_A$}
\Text(0,-10)[l]{$p_B$}
\Text(100,-10)[r]{$p_{B^\prime}$}
\Text(105,70)[l]{$k_1,\, i_1$}
\Text(105,30)[l]{$-k_2,\, i_2$}

\ArrowLine(150,0)(200,0)
\ArrowLine(200,0)(250,0)
\ArrowLine(180,50)(250,70)
\ArrowLine(250,30)(180,50)
\Photon(150,50)(177,50){3}{4}
\ArrowLine(177,50)(180,50)
\Gluon(200,0)(200,55.71){2}{9}
\Gluon(230,0)(230,35.71){2}{6}

\ArrowLine(300,0)(350,0)
\ArrowLine(350,0)(400,0)
\ArrowLine(330,50)(400,70)
\ArrowLine(400,30)(330,50)
\Photon(300,50)(327,50){3}{4}
\ArrowLine(327,50)(330,50)
\Gluon(350,0)(380,35.71){2}{6}
\Gluon(380,0)(350,44.29){2}{8}

\Text(50,-20)[c]{$(1)$}
\Text(200,-20)[c]{$(2)$}
\Text(350,-20)[c]{$(3)$}

\end{picture}
\begin{picture}(400,80)(0,-20)

\ArrowLine(0,0)(50,0)
\ArrowLine(50,0)(100,0)
\ArrowLine(30,50)(100,70)
\ArrowLine(100,30)(30,50)
\Photon(0,50)(27,50){3}{4}
\ArrowLine(27,50)(30,50)
\Gluon(50,0)(80,35.71){2}{6}
\Gluon(80,0)(50,55.71){2}{10}

\ArrowLine(150,0)(200,0)
\ArrowLine(200,0)(250,0)
\ArrowLine(180,50)(250,70)
\ArrowLine(250,30)(180,50)
\Photon(150,50)(177,50){3}{4}
\ArrowLine(177,50)(180,50)
\Gluon(200,0)(200,55.71){2}{9}
\Gluon(230,0)(230,64.29){2}{10}

\ArrowLine(300,0)(350,0)
\ArrowLine(350,0)(400,0)
\ArrowLine(330,50)(400,70)
\ArrowLine(400,30)(330,50)
\Photon(300,50)(327,50){3}{4}
\ArrowLine(327,50)(330,50)
\Gluon(350,0)(350,44.29){2}{7}
\Gluon(380,0)(380,35.71){2}{6}

\Text(50,-20)[c]{$(4)$}
\Text(200,-20)[c]{$(5)$}
\Text(350,-20)[c]{$(6)$}

\end{picture}
\caption[]{The two-gluon exchange one-loop diagrams contributing  to
the ${\cal A}_{Q\gamma^* \rightarrow Qq\bar q}$.}
\end{center}
\end{figure}
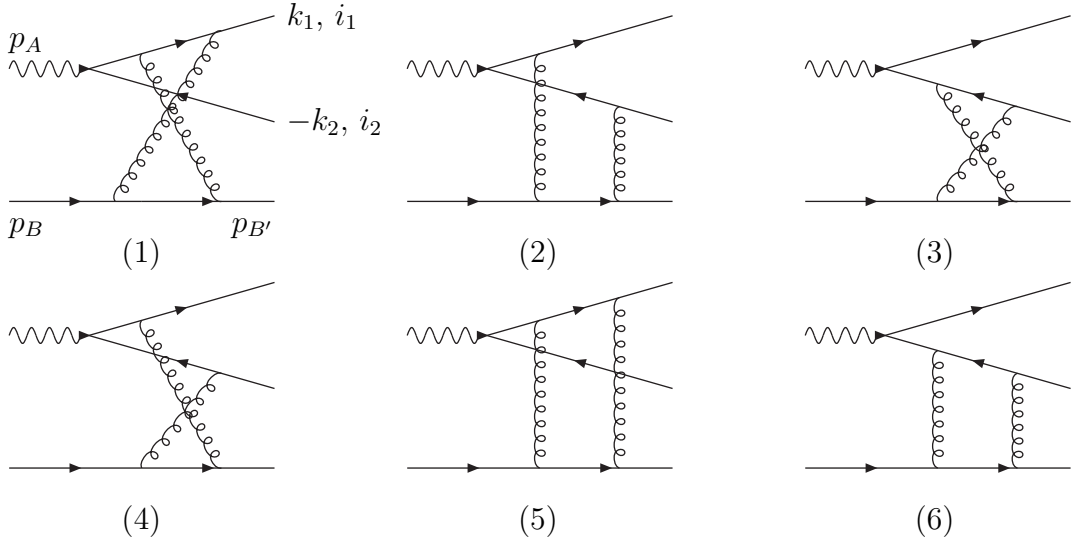
with the following replacement for the colour factors of their lowest lines, in order to project on colour octet
and negative
signature in the $t$-channel
\begin{equation}\label{4}
\left( t^bt^a \right)_{B^\prime B} \rightarrow \frac{1}{2}\left( t^bt^a - t^at^b \right)_{B^\prime B}
= \frac{1}{2}T^c_{ab}t^c_{B^\prime B}.
\end{equation}
Fortunately, to obtain the correction $\Gamma_{\gamma^*q\bar q}^{(2g)c(1)}$ it is enough to calculate only
two of
the diagrams of Fig.1 and the correction $\Gamma_{QQ}^{(2g)c(1)}$ (see Eq.(\ref{3})), which can be found from
the
two-gluon contribution to the quark-quark scattering amplitude, for example. We obtain
\begin{equation}\label{5}
{\cal A}_{Q\gamma^* \rightarrow Qq\bar q}^{(2g)(8,-)(1)} = \frac{1}{4}Nt^c_{i_1i_2}t^c_{B^\prime B}
\left\{ \left[ \left( D_1 + D_2 \right) - \left( 1 \leftrightarrow 2 \right) \right] - \left[ s \leftrightarrow -s
\right] \right\},
\end{equation}
\begin{equation}\label{6}
\Gamma_{QQ}^{(2g)c(1)} = \frac{1}{2}\omega^{(1)}(t)\left[ \frac{1}{\epsilon} + \psi(1) + \psi(1-\epsilon) -
2\psi(1+\epsilon) \right]\Gamma_{QQ}^{c(0)},
\end{equation}
where $\epsilon = (D-4)/2$, $D$ is the space-time dimension,
$D_1$ and $2D_2$ in Eq.(\ref{5}) are the contributions of the  diagrams of Fig.1(1) and Fig.1(2)
with omitted colour generators in their vertices,   and
$1 \leftrightarrow 2$ is the replacement quark $\leftrightarrow$ antiquark.
The one-gluon correction $\Gamma_{\gamma^*q\bar q}^{(1g)c(1)}$ is given by
the diagrams of Fig.2
\begin{figure}[tb]
\begin{center}
\begin{picture}(460,80)(10,-60)

\ArrowLine(40,50)(110,70)
\ArrowLine(110,30)(40,50)
\Photon(10,50)(37,50){3}{4}
\ArrowLine(37,50)(40,50)
\Gluon(80,38.57)(80,3){3}{4}
\ArrowLine(80,3)(80,0)
\Gluon(70,41.43)(70,58.57){3}{2}

\Text(10,60)[l]{$p_A$}
\Text(90,0)[l]{$q,\, \mu,\, c$}
\Text(115,70)[l]{$k_1,\, i_1$}
\Text(115,30)[l]{$-k_2,\, i_2$}

\ArrowLine(170,50)(240,70)
\ArrowLine(240,30)(170,50)
\Photon(140,50)(167,50){3}{4}
\ArrowLine(167,50)(170,50)
\Gluon(210,38.57)(210,3){3}{4}
\ArrowLine(210,3)(210,0)
\GlueArc(190,44.28)(10,165,345){3}{4}

\ArrowLine(280,50)(350,70)
\ArrowLine(350,30)(280,50)
\Photon(250,50)(277,50){3}{4}
\ArrowLine(277,50)(280,50)
\Gluon(320,38.57)(320,3){3}{4}
\ArrowLine(320,3)(320,0)
\GlueArc(320,38.57)(10,-15,165){3}{4}

\ArrowLine(390,50)(460,70)
\ArrowLine(460,30)(390,50)
\Photon(360,50)(387,50){3}{4}
\ArrowLine(387,50)(390,50)
\Gluon(415,43)(415,3){3}{5}
\ArrowLine(415,3)(415,0)
\Gluon(440,35.71)(440,64.29){3}{3}

\Text(60,-20)[c]{$(1)$}
\Text(190,-20)[c]{$(2)$}
\Text(300,-20)[c]{$(3)$}
\Text(410,-20)[c]{$(4)$}

\end{picture}
\begin{picture}(460,80)(10,-40)

\ArrowLine(40,50)(110,70)
\ArrowLine(110,30)(40,50)
\Photon(10,50)(37,50){3}{4}
\ArrowLine(37,50)(40,50)
\Gluon(70,20)(70,3){3}{2}
\ArrowLine(70,3)(70,0)
\Gluon(70,20)(52,44.44){3}{4}
\ArrowLine(52,44.44)(50,47.14)
\Gluon(70,20)(97,31.56){3}{3}
\ArrowLine(97,31.56)(100,32.86)

\Text(75,0)[l]{$\mu,\, q$}
\Text(92,23)[l]{$\nu,\, k$}
\Text(57,28)[r]{$\lambda,\, -q-k$}

\ArrowLine(170,50)(240,70)
\ArrowLine(240,30)(170,50)
\Photon(140,50)(167,50){3}{4}
\ArrowLine(167,50)(170,50)
\Gluon(200,20)(200,3){3}{2}
\ArrowLine(200,3)(200,0)
\Gluon(200,20)(182,44.44){3}{4}
\ArrowLine(182,44.44)(180,47.14)
\Gluon(200,20)(228.50,64.79){3}{7}
\ArrowLine(228.50,64.79)(230,67.14)

\Text(205,0)[l]{$\mu,\, q$}
\Text(222,47)[l]{$\nu,\, k$}
\Text(187,28)[r]{$\lambda,\, -q-k$}

\ArrowLine(280,50)(350,70)
\ArrowLine(350,30)(280,50)
\Photon(250,50)(277,50){3}{4}
\ArrowLine(277,50)(280,50)
\Gluon(320,61.43)(320,3){3}{7}
\ArrowLine(320,3)(320,0)
\Gluon(310,41.43)(310,58.57){3}{2}

\ArrowLine(390,50)(460,70)
\ArrowLine(460,30)(390,50)
\Photon(360,50)(387,50){3}{4}
\ArrowLine(387,50)(390,50)
\Gluon(430,61.43)(430,3){3}{7}
\ArrowLine(430,3)(430,0)
\GlueArc(410,55.72)(10,15,195){3}{4}

\Text(60,-20)[c]{$(5)$}
\Text(190,-20)[c]{$(6)$}
\Text(300,-20)[c]{$(7)$}
\Text(410,-20)[c]{$(8)$}

\end{picture}
\begin{picture}(350,80)(10,-20)

\ArrowLine(40,50)(110,70)
\ArrowLine(110,30)(40,50)
\Photon(10,50)(37,50){3}{4}
\ArrowLine(37,50)(40,50)
\Gluon(80,61.43)(80,3){3}{7}
\ArrowLine(80,3)(80,0)
\GlueArc(80,61.43)(10,15,195){3}{4}

\ArrowLine(170,50)(240,70)
\ArrowLine(240,30)(170,50)
\Photon(140,50)(167,50){3}{4}
\ArrowLine(167,50)(170,50)
\Gluon(195,57)(195,3){3}{7}
\ArrowLine(195,3)(195,0)
\Gluon(220,35.71)(220,64.29){3}{3}

\ArrowLine(280,50)(350,70)
\ArrowLine(350,30)(280,50)
\Photon(250,50)(277,50){3}{4}
\ArrowLine(277,50)(280,50)
\Gluon(310,20)(310,3){3}{2}
\ArrowLine(310,3)(310,0)
\Gluon(310,20)(300,56){3}{5}
\Gluon(310,20)(340,67.14){3}{7}

\Text(60,-20)[c]{$(9)$}
\Text(190,-20)[c]{$(10)$}
\Text(300,-20)[c]{$(11)$}

\end{picture}
\end{center}
\caption[]{The diagrams corresponding to the correction
$\Gamma_{\gamma^*q\bar q}^{(1g)c(1)}$.}
\end{figure}
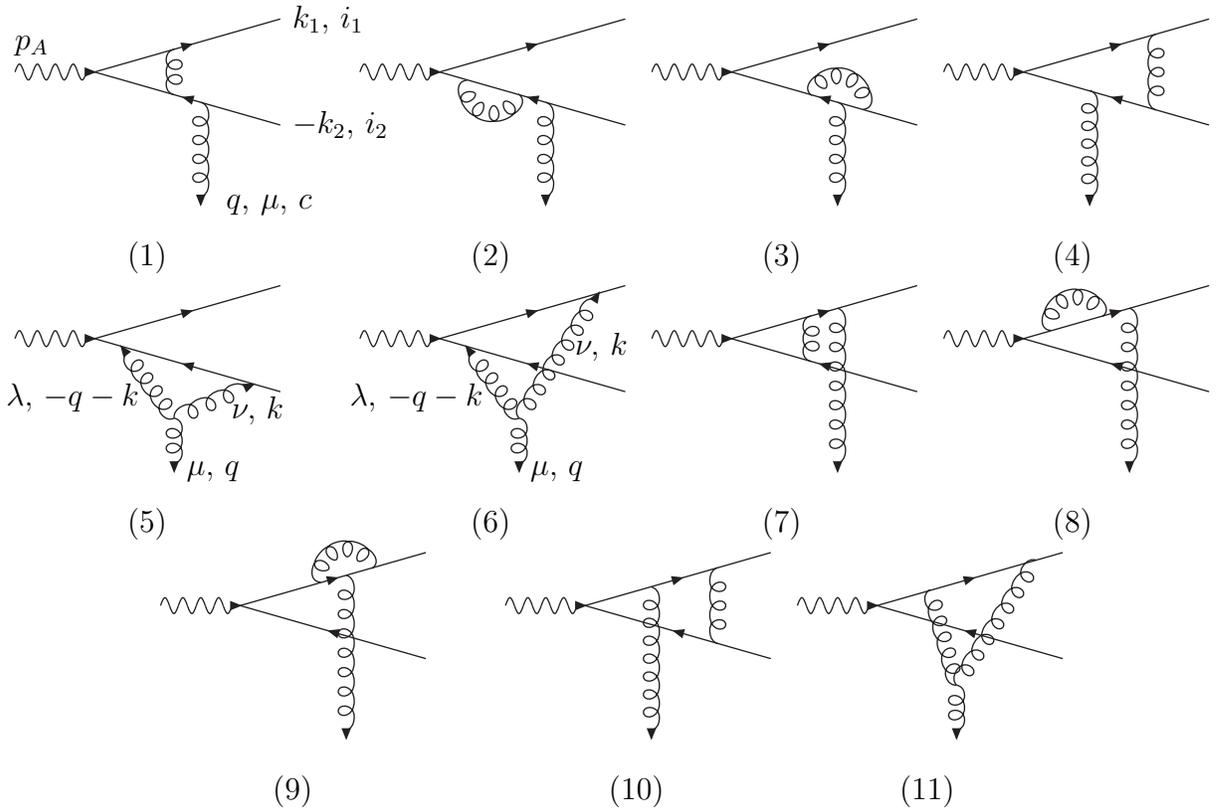
with $ip_2^\mu/s$ instead of the gluon polarization vector,
and again the number of diagrams to be calculated is  reduced due to the
$1 \leftrightarrow 2$ symmetry:
\begin{equation}\label{7}
\Gamma_{\gamma^*q\bar q}^{(1g)c(1)} = Nt^c_{i_1i_2}\left\{ \biggl[ -\frac{2C_F}{N}\left( R_1 + R_2 \right) +
\frac
{N - 2C_F}{N}\left( R_3 + R_4 \right) + R_5 + \tilde R_6 \biggr] - \biggl[ 1 \leftrightarrow 2 \biggr] \right\},
\end{equation}
where the notations $-2R_1, ... , -2R_4, 2R_5$ and $-4\tilde R_6$ were used
for the diagrams of Figs.2(1), ... , (4), (5)
and (6) respectively, with the omitted colour generators in all vertices and
the gluon polarization vector
equal to $ip_2^\mu/s$. Let us note that, whereas the definition of
$R_1, ... , R_4$ is unambiguous, $R_5$ and
$\tilde R_6$ are not yet well defined  because of the presence of
three-gluon vertices in the corresponding diagrams Fig.2(5) and (6). To complete
their definition we show explicitly
 at these diagrams
momenta and vector indices for the three-gluon vertices  for which the
following expression must  be used:
\begin{equation}\label{8}
ig\left[ -g_{\lambda\nu}(2k+q)_\mu + g_{\lambda\mu}(k+2q)_\nu + g_{\nu\mu}(k-q)_\lambda \right].
\end{equation}
As for the last (self energy) contribution $\Gamma_{\gamma^*q\bar q}^{(se)(1)}$,
 it is presented schematically by the diagrams
\begin{figure}[tb]
\begin{center}
\begin{picture}(150,80)(0,-20)

\ArrowLine(30,50)(100,70)
\ArrowLine(100,30)(30,50)
\Photon(0,50)(27,50){3}{4}
\ArrowLine(27,50)(30,50)
\GCirc(40,50){10}{0.5}
\Gluon(40,40)(40,20){3}{2}
\GCirc(40,10){10}{0.5}
\Gluon(40,0)(40,-17){3}{2}
\ArrowLine(40,-17)(40,-20)

\Text(0,60)[l]{$p_A$}
\Text(105,70)[l]{$k_1,\, i_1$}
\Text(105,30)[l]{$-k_2,\, i_2$}
\Text(45,-20)[l]{$q,\, \mu,\, c$}

\end{picture}
\end{center}
\caption[]{The diagrams corresponding to the correction
$\Gamma_{\gamma^*q\bar q}^{(se)c(1)}$.}
\end{figure}
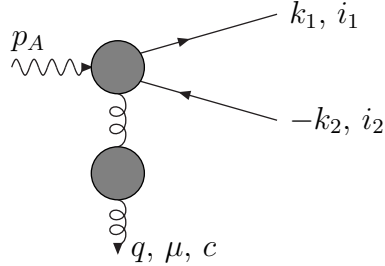
of Fig.3 multiplied with $ip_2^\mu/(2s)$ and can be calculated without any
difficulties: one has  only to know the Born
effective vertex and the one-loop gluon vacuum polarization in the Feynman gauge (we used everywhere the Feynman
gauge
for the gluon fields).

The calculation of all the listed diagrams is quite straightforward but, particularly in the case of diagrams $R_4$ and
$\tilde R_6$ from Eq.(\ref{7}), is rather long and does not give a simple short answer. For these two most
complicated
diagrams we use helicity representation for their regular parts (finite in the physical limit $\epsilon \rightarrow 0$)
$$
\hat\rho \equiv (v_2\bar u_1) = \frac{1}{\sqrt{x(1-x)}}\frac{1}{4}\biggl[ \biggl( (1-x)k_1 + xk_2 - \kappa\frac{p_2}{s}
\biggr)^\mu - 2i\xi e^{\mu\nu\lambda\rho}{k_2}_\nu{k_1}_\lambda\frac{{p_2}_\rho}
{s} \biggr]\gamma_\mu(1 -
\xi\gamma_5),
$$
\begin{equation}\label{9}
e^\mu(\lambda) = \frac{1}{\sqrt{-2t}}\biggl[ (\delta_{\lambda , 1} + \delta_{\lambda , -1})\left( q_\perp^\mu + 2i
\lambda e^{\mu\nu\lambda\rho}q_\nu{p_1}_\lambda\frac{{p_2}_\rho}{s} \right) + \delta_{\lambda , 0}\sqrt
{-2tQ^2}\frac{2p_2^\mu}{s} \biggr],\ \ \lambda = 0, \pm 1,
\end{equation}
where $u_1$ and $v_2$ are bispinor wave functions of the produced quark and antiquark correspondingly,
$\kappa$ is the squared invariant mass of the produced pair and
$\xi = \pm 1$ is the doubled helicity of the quark. For the singular parts of these two diagrams, as well as for all other
diagrams,
we adopt usual spinor representation to present our results. The regular parts
are (we present the result for
$R_6$ which is defined so that $2\tilde R_6 = R_6 - R_6(1 \leftrightarrow 2)$):
$$
R_4^{(r)} = \frac{eq_fg^3}{(4\pi)^2}\frac{2}{\sqrt{2x(1-x)\vec q^{~2}}}\times
$$
$$
\int_0^1\!\!\int_0^1\!\!\int_0^1\frac{dy_1dy_2dy_3y_3}{\left[ -(1-y_1)y_1y_2y_3t - (1-y_1)(1-y_2)y_3t_1 + (1-y_3)
\left( (1-y_1)Q^2 + y_1(-\kappa - i\delta) \right) \right]^2}
$$
$$
\times\biggl\{ \biggl[ -(1-y_1)y_1y_2y_3t - (1-y_1)(1-y_2)y_3t_1 + (1-y_3)\left( (1-y_1)Q^2 - y_1\kappa \right)
\biggr]\biggl[ \left( (1-y_2y_3) \right.
$$
$$
\left.\times(\vec k_1\vec q+i\lambda P) - y_1y_2y_3x\vec q^{~2} \right)(1-x)\delta_{\lambda, -\xi} - y_3\left(
x\vec q^{~2} + \vec k_1\vec q+i\lambda P + (1-y_1)y_2(1-x)\vec q^{~2} \right)
$$
$$
\times x\delta_{\lambda, \xi} \biggr] + (1-y_3)(1-x)\kappa\biggl[ x\left( \sqrt{2}qQx\delta_{\lambda, 0} -
(\vec k_1\vec q+i\lambda P)(\delta_{\lambda, \xi} + \delta_{\lambda, -\xi}) \right) + (1-y_2)\left( (\vec k_2
\vec q-i\lambda P) \right.
$$
$$
\left.\times x\delta_{\lambda, \xi} + (\vec k_1\vec q+i\lambda P)(1-x)\delta_{\lambda, -\xi} -  \sqrt{2}qQx(1-x)
\delta_{\lambda, 0} \right) - y_1x\vec q^{~2}\delta_{\lambda, -\xi} \biggr] + (1-y_1)y_3\biggl[ (1-y_2)
$$
$$
\times\left( \left( (1-x)\vec q^{~4} + ((\vec k_1-\vec k_2)\vec q)(\vec k_2\vec q-i\lambda P) \right)x\delta_{\lambda,
\xi} + (1-x)\vec k_1^{~2}\vec q^{~2}(\delta_{\lambda, -\xi} - \delta_{\lambda, \xi}) \right) + y_1y_2x\vec q^{~2}
$$
$$
\times\left( (1-x)\vec q^{~2}\delta_{\lambda, -\xi} - (\vec k_2\vec q-i\lambda P)\delta_{\lambda, \xi} \right)
\biggr] + (1-y_1)\left( 1-y_3+(1-y_2)y_3(1-x) \right)
$$
$$
\times\biggl[ \left( x(\vec k_2\vec q-i\xi P) - (1-y_3)(\vec k_1\vec q-i\xi P - xt_1) \right)\sqrt{2}qQ
\delta_{\lambda, 0} + (1-y_2)y_3\left( \left( \vec k_1^{~2}\vec q^{~2} - xt_1\right.\right.
$$
$$
\left.\times(\vec k_1\vec q+i\lambda P) \right)(\delta_{\lambda, -\xi}  - \delta_{\lambda, \xi}) + 2(\vec k_1
\vec k_2 - x(1-x)Q^2)(\vec k_1\vec q+i\lambda P)\delta_{\lambda, -\xi}
$$
$$
\left. - (\vec k_1\vec q-i\xi P - xt_1)\sqrt{2}qQ(1-x)\delta_{\lambda, 0} \right)
$$
\begin{equation}\label{10}
+ y_1y_2y_3\vec q^{~2}\left( (\vec k_1\vec q + i\lambda P - xt_1)(\delta_{\lambda, -\xi} - \delta_{\lambda, \xi})
- 2(\vec k_1\vec k_2 - x(1-x)Q^2)\delta_{\lambda, \xi} \right) \biggr] \biggr\},
\end{equation}
$$
R_6^{(r)} = \frac{eq_fg^3}{(4\pi)^2\sqrt{2x(1-x)\vec q^{~2}}}\int_0^1\int_0^1\!\!\!\frac{dy_1dy_2}{\left[
(1-y_2)\left( -(1-y_1)t-y_1t_2 \right) + y_2\left( -(1-y_1)t_1+y_1Q^2 \right) \right]^2}
$$
$$
\times\biggl( y_1\delta_{\lambda, 0}\sqrt{2}qQx\biggl\{ \left( x(1-x)Q^2 - \vec k_1\vec k_2 - i\xi P \right)\left(
1-3(1-x) \right) + 2(1-x)(t_2-t) \biggr\} + (1-y_1)
$$
$$
\times\biggl\{ \biggl[ 2\left( (1-y_2)\left( -(1-y_1)t-y_1t_2 \right) + y_2\left( -(1-y_1)t_1+y_1Q^2 \right) \right)
+ y_1(t_2-t-\kappa) \biggr]x(1-x)
$$
$$
\times\biggl( (\delta_{\lambda, -\xi} + \delta_{\lambda, \xi})(\vec k_1\vec q + i\lambda P) - \delta_{\lambda, 0}
\sqrt{2}qQx \biggr) + \biggl[ 2xt_1 + 2(1-x)t_2 - 2t - y_2(1-x)(t_2-t-\kappa) \biggr]
$$
$$
\times\biggl( \delta_{\lambda, 0}\sqrt{2}qQx(1-x) -  \delta_{\lambda, -\xi}(1-x)(\vec k_1\vec q + i\lambda P) -
\delta_{\lambda, \xi}x(\vec k_2\vec q - i\lambda P) \biggr)
$$
$$
- \biggl[ t\left( 2x\delta_{\lambda, \xi} + 3(1-x)\delta_{\lambda, -\xi} \right) + y_1x\left( t\delta_{\lambda, -\xi}
- \delta_{\lambda, 0}\sqrt{2}qQx \right) \biggr]\left( x(1-x)Q^2 - \vec k_1\vec k_2 - i\xi P \right)
$$
$$
+ \biggl[ 3\left( (1-x)^2t_2\delta_{\lambda, -\xi} - x^2t_1\delta_{\lambda, \xi} \right) + y_1x(1-x)\left( t_2\delta
_{\lambda, -\xi} - t_1\delta_{\lambda, \xi} \right) - y_1xQ^2\delta_{\lambda, \xi} \biggr](\vec k_1\vec q + i\lambda P)
$$
\begin{equation}\label{11}
+ xt\biggl[ 3(1-x)^2Q^2\delta_{\lambda, -\xi} - 3\vec k_1^{~2}\delta_{\lambda, \xi} + y_1xQ^2\left( \delta
_{\lambda, -\xi}(1-x) + \delta_{\lambda, \xi}x \right) + 2y_2(1-x)\kappa\delta_{\lambda, \xi} \biggr] \biggr\} \biggr),
\end{equation}
with the following notations
$$
t_1 = (p_A - k_1)^2 = -\frac{\vec k_1^{~2} + x(1-x)Q^2}{x},\ \ t_2 = (p_A - k_2)^2 = -\frac{\vec k_2^{~2} +
x(1-x)Q^2}{1 - x},
$$
\begin{equation}\label{12}
P = 2e^{\mu\nu\lambda\rho}{k_1}_\mu{k_2}_\nu{p_1}_\lambda\frac{{p_2}_\rho}{s},\ \ P^2 = \vec k_1^{~2}\vec
k_2^{~2} - (\vec k_1\vec k_2)^2.
\end{equation}
We present the one-loop correction to the vertex
$\Gamma_{\gamma^*q\bar q}^{c}$ in the form
\begin{equation}\label{13}
\Gamma_{\gamma^*q\bar q}^{c(1)} = \Gamma_{\gamma^*q\bar q}^{(sing)c(1)} + \Gamma_{\gamma^*q\bar q}
^{(reg)c(1)},\ \ \Gamma_{\gamma^*q\bar q}^{(reg)c(1)} = Nt^c_{i_1i_2}\biggl\{ \biggl[ \frac{N - 2C_F}{N}R_4^{(r)}
+ R_6^{(r)} \biggr] - \biggl[ 1 \leftrightarrow 2 \biggr] \biggr\},
\end{equation}
with the contributions of all other discussed diagrams, as well as the singular parts of $R_4$ and $\tilde R_6$, included
into
$\Gamma_{\gamma^*q\bar q}^{(sing)(1)}$:
$$
\Gamma_{\gamma^*q\bar q}^{(sing)c(1)}\left( eq_fg^3Nt^c_{i_1i_2}\frac{\Gamma(2-\epsilon)}{(4\pi)^{2+\epsilon}}
\frac{1}{2\epsilon} \right)^{-1} = \biggl[ \bar u_1\biggl( \left( -t \right)^\epsilon\biggl\{ \frac{5}{3} - \frac{2}{3}
\frac{n_f}{N} + 4(1+\epsilon)\ln(1-x) + 2\epsilon\biggl( \frac{2}{9}\frac{n_f}{N}
$$
$$
- \frac{8}{9} - \psi^\prime(1) \biggr) \biggr\}\frac{\hat\Gamma_1}{t_1} + \frac{2C_F}{N}\frac{\hat\Gamma_1}
{\left( -t_1 \right)^{1-\epsilon}} + \frac{2C_F}{N}\int_0^1\frac{dy}{\left( (1-y)Q^2 - yt_1 \right)^{1-\epsilon}}
\biggl\{ (1+2\epsilon)\frac{Q^2}{t_1}\hat \Gamma_1 + 2\epsilon Q
$$
$$
+ y\biggl[ (1-2\epsilon)\left( \frac{Q^2}{t_1} + 1 \right)\hat \Gamma_1 + 2(2-\epsilon)\left( {k_1}_\perp e_\perp + xQ
\right) \biggr] \biggr\} + \frac{1}{N}\int_0^1\frac{dy}{\left( -(1-y)t-yt_1 \right)^{1-\epsilon}}
$$
$$
\times\biggl\{ 2\left( (1+3\epsilon)N - (1+2\epsilon)C_F \right)\frac{t}{t_1}\hat\Gamma_1 - (2+\epsilon)N\hat
\Gamma_1 + (1+2\epsilon)N(1-x)\left( \hat\Gamma_1 + \hat\Gamma_2 - 2q_\perp e_\perp \right)
$$
$$
- y^\epsilon4(1+\epsilon)N\left( \frac{t}{t_1} - 1 \right)\hat\Gamma_1 + y\biggl[ 2\left( (1-\epsilon)N - (1-2\epsilon)
C_F \right)\left( \frac{t}{t_1} - 1 \right)\hat\Gamma_1 + \left( (1-2\epsilon)N \right.
$$
$$
\left. - 2(2-\epsilon)C_F \right)(1-x)\left( \hat\Gamma_1 + \hat\Gamma_2 - 2q_\perp e_\perp \right) \biggr] \biggr\}
+ \frac{N-2C_F}{N}\int_0^1\int_0^1\int_0^1dy_1dy_2dy_3y_2^{\epsilon-1}y_3^{\epsilon+1}2\epsilon\kappa
$$
$$
\times\frac{y_2(1-y_2)(1-x)\left( \hat \Gamma_1 + \hat \Gamma_2 - 2q_\perp e_\perp \right) - (1-y_2)\hat
\Gamma_1 - 2(1-y_3)\left( {k_1}_\perp e_\perp + xQ \right)}{\left[ -(1-y_1)y_1y_2y_3t - (1-y_1)(1-y_2)y_3t_1
+ (1-y_3)\left( (1-y_1)Q^2 + y_1(-\kappa - i\delta) \right) \right]^{2-\epsilon}}
$$
$$
+ \int_0^1\int_0^1\frac{dy_1dy_2}{\left[ (1-y_2)\left( -(1-y_1)t-y_1t_2 \right) + y_2\left( -(1-y_1)t_1+y_1Q^2
\right) \right]^{2-\epsilon}}\biggl\{ y_1^{\epsilon-1}(1-y_1)y_2^{-\epsilon}
$$
$$
\times\left( x^\epsilon(1-x)^{-\epsilon} - 2\epsilon^2\psi^\prime(1) \right)2t\hat \Gamma_1 + \left( y_1^\epsilon
y_2^{-\epsilon}x^\epsilon(1-x)^{-\epsilon} - 1 \right)4(1-x)tQ + (1-y_1)\biggl[ 2(t_2-t)\hat\Gamma_1
$$
$$
- xt_1\left( \hat \Gamma_1 + \hat \Gamma_2 - 2q_\perp e_\perp \right) + 4t({k_1}_\perp e_\perp + xQ) \biggr]
+ y_1(1-y_1)\biggl[ 4(t_2 -t)({k_1}_\perp e_\perp + xQ)
$$
\begin{equation}\label{14}
+ x(t_1+Q^2)\left( \hat\Gamma_1 + \hat\Gamma_2 - 2q_\perp e_\perp \right) \biggr] \biggr\} \biggr)\frac{\not p_2}
{s}v_2 \biggr] - \biggl[ 1 \leftrightarrow 2 \biggr],
\end{equation}
\begin{equation}\label{15}
\hat\Gamma_1 = \frac{{\not k_1}_\perp\not e_\perp - 2x({k_1}_\perp e_\perp) + 2x(1-x)Q}{x},\ \ \hat\Gamma_2
= -\frac{{\not k_2}_\perp\not e_\perp - 2x({k_2}_\perp e_\perp) - 2x(1-x)Q}{1-x}.
\end{equation}
The result is valid with the accuracy up to non-vanishing in the limit $\epsilon \rightarrow 0$ terms, which is enough
for the subsequent
calculation of the impact factor. Although all the integrals can be calculated as expansions in $\epsilon$ with the
necessary accuracy (even if such the calculation is very long, it is quite straightforward), we present our result
in the integral form to have the possibility to use standard Feynman parametrization and to change orders of
integration over
all Feynman parameters in the calculation of the impact factor.
The details of our calculation with more
discussion of physics will be given elsewhere \cite{1}. More details concerning
the importance of the impact factors of colourless particles
 for the BFKL approach and the connection between Reggeon effective interaction vertices and impact
factors can be found in \cite{2}.

\end{document}